\documentclass[prd,preprint,superscriptaddress,a4paper,showpacs,showkeys,11pt,nofootinbib]{revtex4-1}

\usepackage{graphicx}
\usepackage{dcolumn}
\usepackage{bm}
\usepackage{multirow}
\usepackage{tabularx}
\usepackage{hyperref}
\usepackage{commath}
\usepackage{amsmath}
\usepackage{amssymb}

\begin{document}

\title{Non-exclusive particle production by $\gamma \gamma$ interactions in $pp$ collisions at the LHC}


\author{G. Gil da Silveira}

\email[]{gustavo.silveira@cern.ch}

\affiliation{Instituto de F\'{\i}sica, Universidade Federal do Rio Grande do Sul\\
Caixa Postal 15051, CEP 91501-970, Porto Alegre, RS, Brazil}

\affiliation{Departamento de F\'{\i}sica Nuclear e de Altas Energias, Universidade do Estado do Rio de Janeiro\\
CEP 20550-013, Rio de Janeiro, RJ, Brazil}


\author{V. P. Gon\c calves}

\email[]{barros@ufpel.edu.br}

\affiliation{Instituto de F\'{\i}sica e Matem\'atica, Universidade Federal de
Pelotas\\
Caixa Postal 354, CEP 96010-090, Pelotas, RS, Brazil}


\author{G. G. Vargas Veronez}


\affiliation{Instituto de F\'{\i}sica e Matem\'atica, Universidade Federal de Pelotas\\
Caixa Postal 354, CEP 96010-090, Pelotas, RS, Brazil}


\begin{abstract}
Particle production in two-photon interactions at hadronic collisions is becoming increasingly relevant in the LHC physics programme as a way to improve our understanding of the Standard Model and search for signals of New Physics. A key ingredient for the study of these interactions in $pp$ collisions is the description of the photon content of the proton, which allow us to derive predictions for the cross sections associated to events where occur the proton dissociation (non - exclusive processes)  and for those where both incident protons remain intact (exclusive processes). In this paper, a detailed comparison of the different models for the elastic and inelastic photon distributions found in the literature is presented and the current theoretical uncertainty is estimated. The impact on the invariant mass distribution for the dimuon production is analyzed. Moreover, the relative contribution of non - exclusive events is estimated and its dependence on the invariant mass of the pair is presented. We demonstrate that the predictions for production of pairs with large invariant mass is strongly dependent on the model assumed to describe the elastic and inelastic photon distributions and that the ratio between non - exclusive and exclusive cross sections present a mild energy dependence. Finally, our results indicate that a future experimental analysis of the non - exclusive events will be useful to constrain the photon content of proton.

\end{abstract}




\maketitle


\section{Introduction}

The LHC experiments have focused part of its physics goals into the particle production by two - photon (electromagnetic) interactions in order to improve our understanding of the Standard Model (SM) and search for signals of New Physics \cite{review_forward}. Typical examples are the studies about the production of dileptons and $W^+ W^-$ pairs   
  \cite{Chatrchyan:2011ci,Chatrchyan:2012tv,Aad:2015bwa,Cms:2018het,Aaboud:2017oiq,Aad:2020glb,Chatrchyan:2013foa,Khachatryan:2016mud,Aaboud:2016dkv}, which investigate the exclusive production of pairs with low and high invariant masses,  covering distinct parts of the phase space available at the LHC energies. Exclusive production means that the final state is composed only by the centrally produced pair, with large rapidity gaps with no tracks between the pair, detected by the central detectors, and the beam line direction. Such signature differs from the usual QCD production by the absence of particle (gluon) radiation that populates the detector, destroying the gap and making it very difficult to be observed. Since no photon radiation occurs during the electromagnetic interaction, the experimental signature is a large pseudorapidity gap, $\Delta\eta$, with no energy deposits in the detector. While the exclusive dilepton production by $\gamma \gamma$ interactions  is considered a luminosity monitor \cite{Shamov:2002yi,Khoze:2000db}, the experimental data for the exclusive $W^+ W^-$ production has been used to constrain the magnitude of quartic anomalous couplings. Recently, the detailed studies performed in Refs.~\cite{Goncalves:2020saa,marek_dark} have demonstrated  the potentiality of exclusive processes to probe the $t  \bar{t}$ production as well the SUSY particle production in compressed mass scenarios (For other recent studies see, e.g. Refs.~\cite{Goncalves:2020vvw,Baldenegro:2020qut}). These promising results strongly motivate the improvement of our understanding about the electromagnetic interactions in the LHC.

\begin{figure}[t]
\centering
\includegraphics[width=1.\textwidth]{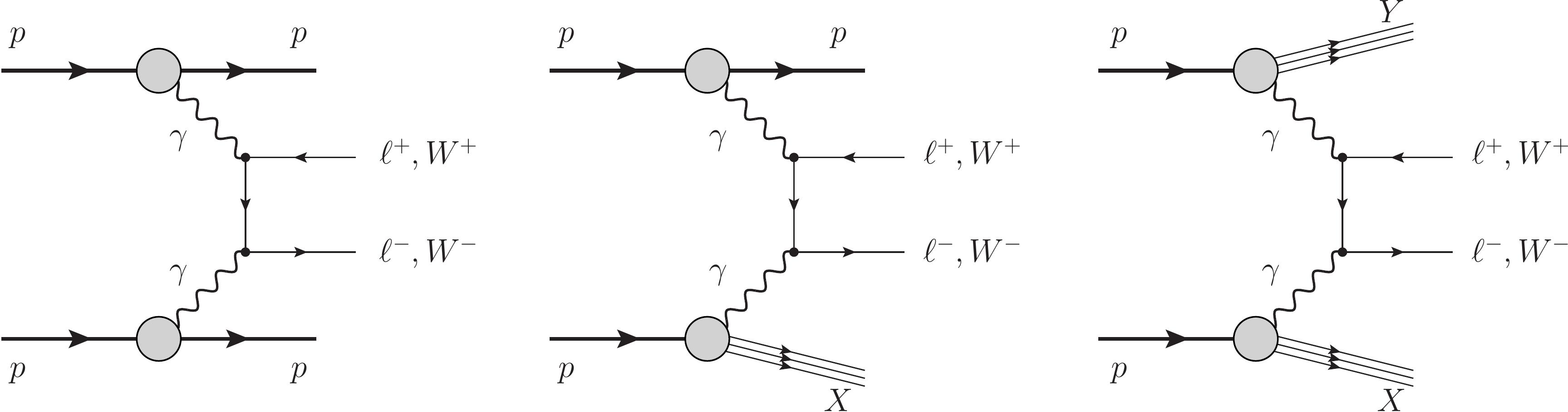}
\caption{\label{Fig:diagram} Particle production  by two-photon interactions in $pp$ collisions through the elastic (left panel),  semi - elastic (central panel) and  inelastic (right panel) processes.}
\end{figure}

Particle production by two - photon interactions may be classified in  elastic, the semi - elastic, and inelastic processes \cite{review_forward}, as illustrated in Fig.~\ref{Fig:diagram}. In the elastic case, both incident protons remain intact, and can be detected in the final state using dedicated forward detectors, as the AFP \cite{afp1,afp2,Cerny:2020rvp} and PPS \cite{pps} in the LHC, which have been installed in association with both ATLAS and CMS detectors, respectively. In contrast, the semi - elastic and inelastic processes are characterized by the dissociation of one or both protons, respectively. While the elastic case is a typical example of an exclusive process, the semi - elastic and inelastic interactions are usually denoted non - exclusive one.  In all cases, two rapidity gaps will be present, due to the photon exchange. Two  current challenges are the theoretical treatment of these different processes and its experimental separation. In order to illustrate the first aspect, let's consider the dimuon production by $\gamma \gamma$ interactions (Similar analysis is valid for other final states). One has that  the associated hadronic cross sections   can be factorized as follows
\begin{eqnarray}
\sigma^i  \propto {\cal{L}}_{eff}^i \times \hat{\sigma}(\gamma \gamma \rightarrow \mu^+ \mu^-),     
\end{eqnarray}
where  ${\cal{L}}^i_{eff}$ is the effective photon-photon luminosity for elastic, semi - inelastic and inelastic processes and the elementary cross section $\hat{\sigma}(\gamma \gamma \rightarrow \mu^+ \mu^-) $ is well - known from Quantum Electrodynamics (QED). Therefore, the main uncertainty in the calculation of the different contributions is associated to the modelling of  ${\cal{L}}^i_{eff}$, which can be  expressed in terms of the elastic and inelastic photon distributions as follows
\begin{eqnarray}
& \,& \,\,\, {\cal{L}}^{\textrm{el}}_{eff} \propto x_{1}f^{\textrm{el}}_{\gamma,1}(x_{1};Q^{2}) x_{2}f^{\textrm{el}}_{\gamma,2}(x_{2};Q^{2}), \\
&\,& \,\,\, {\cal{L}}^{\textrm{semi}}_{eff} \propto x_{1}f^{\textrm{inel}}_{\gamma,1}(x_{1};Q^{2}) x_{2}f^{\textrm{el}}_{\gamma,2}(x_{2};Q^{2}) + x_{1}f^{\textrm{el}}_{\gamma,1}(x_{1};Q^{2}) x_{2}f^{\textrm{inel}}_{\gamma,2}(x_{2};Q^{2}), \\
&\,& \,\,\, {\cal{L}}^{\textrm{inel}}_{eff} \propto x_{1}f^{\textrm{inel}}_{\gamma,1}(x_{1};Q^{2}) x_{2}f^{\textrm{inel}}_{\gamma,2}(x_{2};Q^{2}),
\end{eqnarray}
where $x_i$ are the  momentum fraction of the proton taken by the photons and $Q^2$ is the photon virtuality. The elastic photon distribution $f^{\textrm{el}}$ is associated to the probability that a proton emits a photon and remains intact. Such distribution can be expressed  in terms 
of the  electric and magnetic  form factors  using the Equivalent Photon Approximation (EPA) method \cite{epa,Terazawa} and have been estimated in the literature assuming different approximations, as detailed in the next Section. On the other hand, the inelastic photon distribution $f^{\textrm{inel}}$ provides the probability for a photon emission from a proton in an inelastic interaction and can be estimated assuming that the photon is a constituent of the proton, along with quarks and gluons, with its contribution being  derived by solving the DGLAP evolution equations modified by the inclusion of the QED parton splittings. In recent years, the determination of photon PDF in a global analysis was performed by different groups assuming distinct assumptions  for, e.g.,  the initial conditions and the treatment of the higher order corrections \cite{Manohar:2017eqh,Bertone:2017bme,Harland-Lang:2019pla}. The contribution of the elastic, semi - elastic and inelastic  processes have been estimated in the literature assuming distinct approximations for the calculation of the elastic photon distribution as well as for different choices of the inelastic photon PDF (See, e.g., Refs.~\cite{vicgus,daSilveira:2015hha,antoni_gus,Luszczak:2015aoa,Lebiedowicz:2018muq,Forthomme:2018sxa,Luszczak:2018dfi,nosdilepton_pp}). One of our goals is to estimate the current theoretical 
uncertainty associated with these different modellings of the photon 
distributions and determine its impact on the predictions for the cross sections.

Another goal of this paper is to estimate the 
relative contribution between exclusive and non-exclusive events for the distinct treatments of the photon distributions and determine its dependence on the invariant mass of the diphoton system. Such study is strongly motivated by the  CMS analyzes performed in Refs.~\cite{Chatrchyan:2013foa,Khachatryan:2016mud}, which have presented a data-driven method to account for such relative contribution. In these analyses, the relative contribution of  non-exclusive to elastic events was obtained by counting the measured events and comparing it to the theoretical expectation for the elastic contribution as follows:
\begin{eqnarray}
F = \left. \frac{N_{\mu\mu\textrm{(data)}}-N_{\textrm{DY}}}{N_{\textrm{elastic}}} \right|_{M(\mu^{+}\mu^{-})>160\textrm{ GeV}} 
\end{eqnarray}
where $N_{\mu\mu\textrm{(data)}}$ is the total number of events passing the selection criteria, $N_{\textrm{DY}}$ the total number of events identified as coming from the Drell-Yan production process related to events with one or more extra tracks, and $N_{\textrm{elastic}}$ is the estimated number of elastic events from theory.  The resulting number  was employed to scale up the event samples produced by an event generator for the exclusive production of $W$ pairs, which provides elastic events only, and derive an estimate of the non-exclusive contribution. One of the main assumptions of this method is that the multiplicative factor is a constant, independent on the invariant mass of the final system. In our study we will calculate the elastic, semi - elastic, and inelastic cross sections considering the different models for the elastic and inelastic photon distributions and the ratio between these distinct contributions will be estimated. It will allow us to test the assumption implemented in the CMS analyses as well as to estimate the current theoretical uncertainty on the predictions for the ratio between non - exclusive and exclusive processes at the LHC.

This paper is organized as follows. In the next Section, we will present a brief review of the modelling of elastic and inelastic photon distributions. The different approximations for the elastic distribution, usually found in the literature, will be discussed, and the recent parameterizations for the photon PDF will be presented. A comparison between these different models will be shown.  In Section \ref{sec:res} we will present the associated predictions for the elastic, semi - elastic and inelastic dimuon production and the current theoretical uncertainty on the predictions of the invariant mass distributions  will be estimated. Moreover, the ratio between non - exclusive and exclusive processes will be  estimated for the distinct models for the elastic and inelastic photon distributions. Finally, in Section \ref{sec:sum}, we will summarize our main results and conclusions.



\section{The photon content of the proton}
\label{sec:xsecs}
The general expression for the production cross section of a final state $X$   by two -- photon interactions in $pp$ collisions can be derived assuming the validity of the factorization theorem and is given by \cite{epa,Terazawa}
\begin{eqnarray}
\sigma(pp\to p \otimes X \otimes p) = S_{\gamma}^{2}\int_{0}^{1}\int_{M^2/sx_1}^{1} f_{\gamma,1}(x_{1},Q^2) f_{\gamma,2}(x_{2},Q^2) \hat{\sigma}_{\gamma\gamma\to X} \dif x_{1} \dif x_{2} \,\,,
\end{eqnarray}
where $\otimes$ represents a rapidity gap in the final state and $S_{\gamma}^{2}$ is the so - called survival factor, which takes into account the requirement that there is no hadronic interactions between the incident protons that can produce extra particles and destroy the rapidity gaps. For two -- photon interactions, such corrections are expected to be small and in what follows we will assume that $S_{\gamma}^{2} = 1$ (For recent discussions about the treatment of the survival factor see, e.g., Refs.~\cite{Dyndal:2014yea,Harland-Lang:2020veo}). 
Moreover, $x_i$ are the fractions of the hadron  energies carried by the photons and $Q^2$ has to be identified with a momentum scale of the  process.
As $\hat{\sigma}_{\gamma\gamma\to X}$ is in general well - known, the main ingredient in the analysis of photon - induced processes is the photon distribution associated to the incident protons. While the photon distribution associated to a a charged  pointlike fermion can be precisely determined using the equivalent photon approximation formulated  many years ago by Fermi \cite{Fermi} and developed by Williams \cite{Williams} and Weizs\"acker \cite{Weizsacker}, the calculation of the photon distribution of the hadrons still is a subject of debate, due to the fact that they are not pointlike particles. This case makes necessary to distinguish between the  elastic and inelastic components.  The elastic component, $f^{\textrm{el}}$, can be estimated analysing the transition $p \rightarrow \gamma p$ taking into account the effects of the hadronic form factors, with the hadron remaining intact in the final state \cite{epa,kniehl}. In contrast, the inelastic contribution, $f^{\textrm{inel}}$, is associated to the transition $p \rightarrow \gamma X$, with $X \neq p$, and  can be estimated taking into account the partonic structure of the hadrons, which can be a source of photons (see, e.g., Refs.~\cite{rujula,drees_godbole,pisano,mrstqed,nnpdf,martin_ryskin,Manohar:2016nzj}). In what follows,  the distinct approaches present in literature will be reviewed and a comparison between its predictions is performed.

The elastic photon distribution can be estimated  using the general expression for the equivalent photon flux of an extended object, which is given by \cite{epa}
\begin{eqnarray}
f^{\textrm{el}}_{\gamma}(x,Q^2) = \frac{\alpha Z^2}{\pi} \frac{1 - x + 0.5x^2}{x} \int_{Q^2_{min}}^{Q^2} dq_T^2 \frac{q_T^2 - Q^2_{min} }{q_T^4} |F(q_T^2)|^2 \,\,,
\label{fluxgeral}
\end{eqnarray}
where $q_T^2$ is the transverse momentum of the emitted photon and $F(q_T^2)$ is associated form factor. Moreover, $Q^2_{min} \approx (x m)^2/(1-x)$, with $m$ the 
mass of the proton projectile. 
The presence of the form factor cuts off the photon flux above $q_T \simeq 2$ GeV$^2$, which implies that the elastic photon distributions becomes basically independent of $Q^2$, i.e., $f^{\textrm{el}}_{\gamma}(x,Q^2) \approx f^{\textrm{el}}_{\gamma}(x)$. As a consequence,  the elastic processes, represented in Fig.~\ref{Fig:diagram} (a), can be considered as the interaction between two quasi-real photons.  
Considering only the electric dipole form factor for the proton, $F_E (q_T^2) = 1/(1+q_T^2/0.71\,$GeV$^2)^2$, the following expression for the elastic photon distribution can be obtained 
\begin{eqnarray}
f^{\textrm{el}}_{\gamma} (x) &=&\frac{\alpha}{\pi}\left(\frac{1-x+0.5x^{2}}{x}\right)
\left[\frac{A+3}{A-1}\ln(A)-\frac{17}{6}-\frac{4}{3A}+\frac{1}{6A^{2}}\right] \,\,,
\end{eqnarray}
where $A=1+(0.71 \mathrm{GeV}^{2})/Q_{min}^{2}$. We denote this model by {\it Electric (E)} in what follows. 
If the term containing $Q^2_{min}$ in Eq. (\ref{fluxgeral}) is disregarded, the 
 equivalent photon spectrum of high energy protons
is given as follows
\begin{eqnarray}
f^{\textrm{el}}_{\gamma}(x)&=&\frac{\alpha}{\pi}\left(\frac{1-x+0.5x^{2}}{x}\right)\left[\ln(A)
-\frac{11}{6}+\frac{3}{A}-\frac{3}{2A^{2}}+\frac{1}{3A^{3}}\right]\,\,.
\end{eqnarray}
This expression was derived originally by Dress and Zeppenfeld in Ref. \cite{dz} and will be denoted \emph{DZ} hereafter.
In Ref.~\cite{kniehl}, the author studied the effect of including the magnetic dipole moment and the corresponding magnetic form factor of the proton, obtaining a  more precise expression for the elastic photon distribution, which will be denoted {\it Electric + Magnetic (EM)} hereafter.  
 Finally, another model for $f^{\textrm{el}}_{\gamma}$ found in the literature, is given by 
\begin{eqnarray}
f^{\textrm{el}}_{\gamma}(x) = \frac{\alpha Z^2}{\pi x} \left[ 2 \xi K_0 (\xi) K_1 (\xi) - \xi^2 ( K_1^2 (\xi) - K_0^2 (\xi) ) \right],
\label{Eq:point}
\end{eqnarray}  
where $K_0$ and $K_1$ are modified Bessel functions and $\xi \equiv x m b_{min}$. Such expression is derived using the Weizsacker - Williams approximation assuming a pointlike form factor and that $b_{min}$~=~0.7~fm.

\begin{figure}[t]
\centering
\includegraphics[width=.49\textwidth]{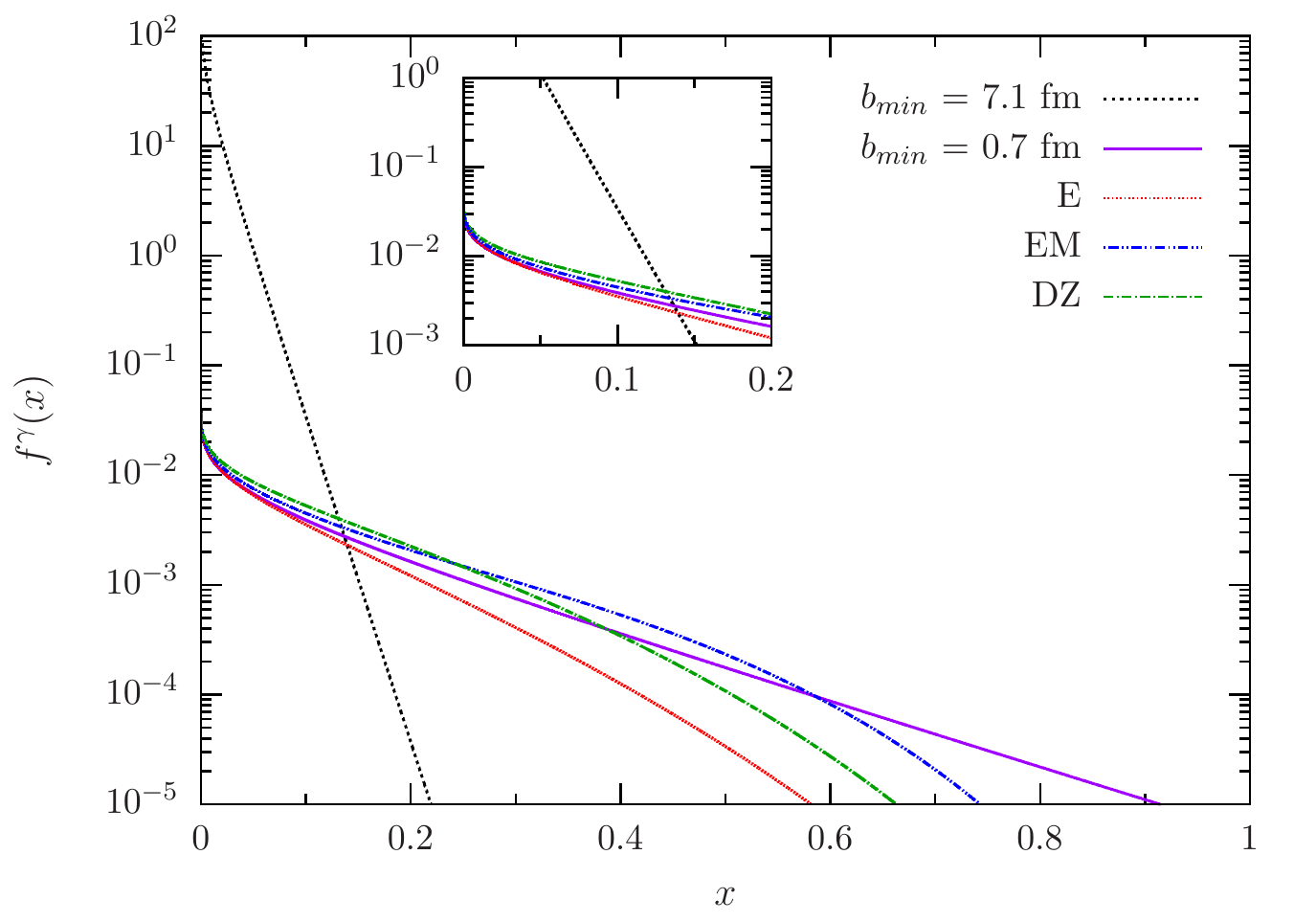}
\includegraphics[width=.49\textwidth]{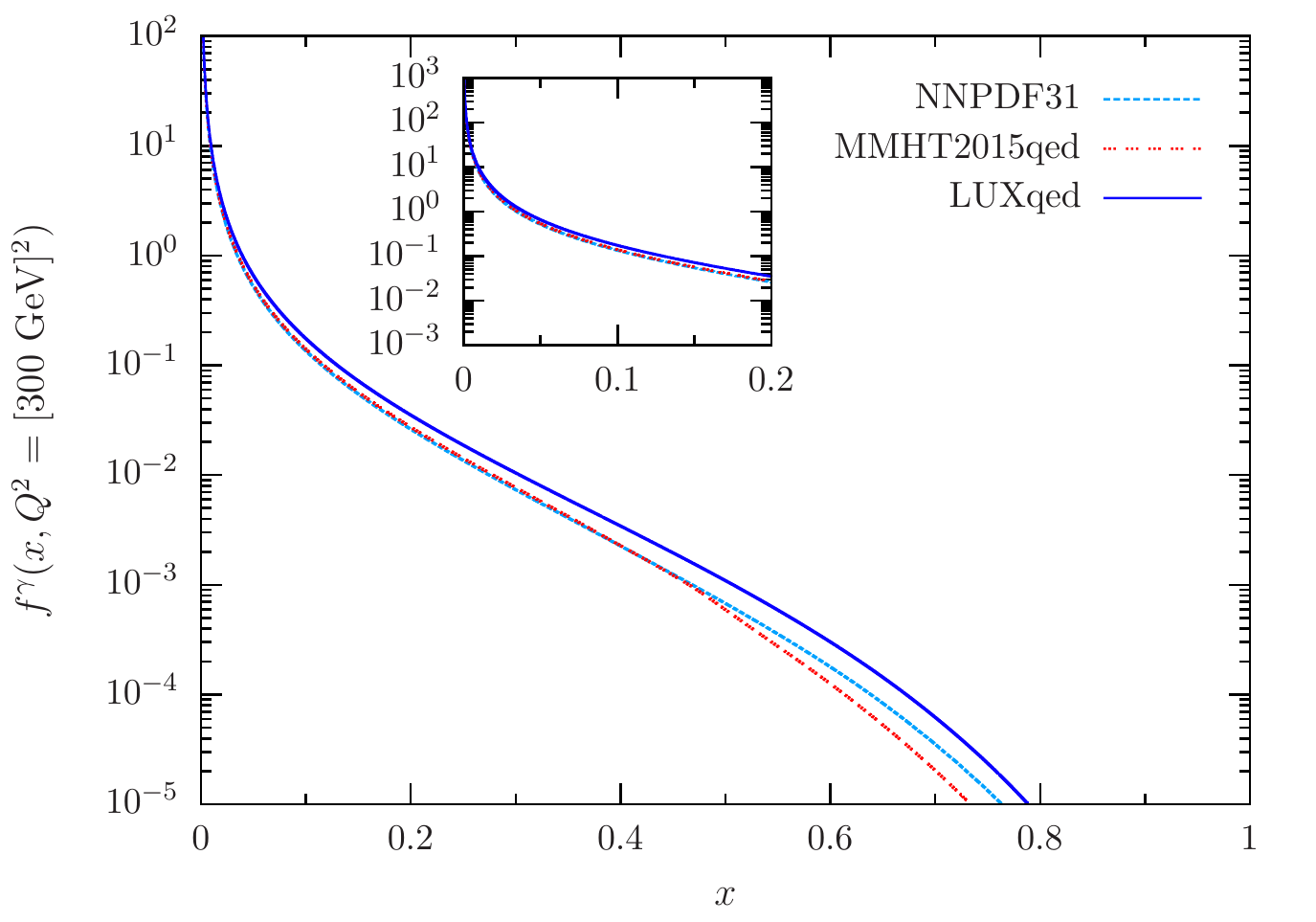}
\caption{\label{Fig:fgamma}  Comparison between the distinct models for the elastic (left panel) and inelastic (right panel) photon distributions of the proton.}
\end{figure}

 In Fig.~\ref{Fig:fgamma} (left panel)
we present a comparison between these different models for the elastic photon distribution for the proton. For completeness, we also present the prediction for the nuclear photon flux, which is derived using Eq. (\ref{Eq:point}) and assuming that $Z = 82$ and $b_{min}$~=~7.1~fm.  A basic characteristic of all models for $f^{\textrm{el}}_{\gamma}$ is that they diminish  with energy approximately like $1/x$.  Consequently, the photon spectrum is strongly peaked at low $x$, implying that the dimuon and $W$ pair cross sections will be dominated by the production of pairs with small invariant mass. Due to the factor $Z^2$ in Eq.~(\ref{Eq:point}), the nuclear photon flux is strongly enhanced in comparison to the proton one. However, it has a steeper decrease at larger values of $x$, which implies that the production of pairs with large invariant masses is suppressed in heavy -- ion collisions in comparison to $pp$ one. The comparison between the distinct models for proton photon flux indicates that its predictions are similar for low $x$ ($\le 0.05$). However, the difference  increases for larger values of  $x$. In particular, for  $x = 0.1$ the difference among the EM and E models is $\approx$ 20 \%, increasing for one order of magnitude for  $x = 0.4$. Finally, for larger values of $x$ the predictions for the elastic photon distributions are strongly model dependent. Such large theoretical uncertainty has a direct impact on the predictions for the elastic production of pairs with large invariant mass, as we will demonstrate in the next Section.

As discussed in the Introduction, in addition to the elastic case, one has to take into account the possibility of proton dissociation during the two - photon interactions, as shown in the central and right panels of Fig.~\ref{Fig:diagram}. The calculation of these contributions is dependent on the inelastic photon distribution   $f^{\textrm{inel}}$. Currently, there is no analytic framework to account for such contribution. Recent studies have considered the photon as a parton inside the proton, with an associated photon PDF, and derived a prediction for this distribution solving the DGLAP evolution equations modified to include the QED contributions.  In this work, we employ a set of the recent parametrizations for the photon PDF: LUXqed17 \cite{Manohar:2017eqh}, MMHT2015qed \cite{Harland-Lang:2019pla}, and NNPDF31luxQED \cite{Bertone:2017bme}. 
All these parametrizations are based on the approach proposed in Ref.~\cite{Manohar:2016nzj} (See also Ref.~\cite{Luszczak:2015aoa}), which have allowed to estimate the photon PDF from the lepton - proton structure functions and have reduced the uncertainty on its determination. However, they differ in the distinct methodologies used to extract the photon PDF. For example, the NNPDF31luxQED parametrization is build upon the NNPDF3.1 fit, with the photon PDF being determined by means of a global PDF analysis supplemented by the theoretical constraint proposed in Ref. \cite{Manohar:2016nzj}. In contrast, the MMHT2015qed parametrization is the result of including QED effects in the MMHT framework. 
It is also important to emphasize that some of these parametrizations does not provide the predictions for $f^{\textrm{inel}}$, but instead for $f^{\textrm{el}} + f^{\textrm{inel}}$, which is usually denoted inclusive photon PDF. For these cases, we must to subtract the elastic component in order to estimate the associated inelastic photon distribution. 
In Fig. \ref{Fig:fgamma} (right panel) we present  a comparison between the results for $f^{\textrm{inel}}$ predicted for these  distinct parameterizations for $Q = 300$ GeV, with the elastic contribution  subtracted from the inclusive photon PDF provided by the  LUXqed and NNPDF31luxQED parametrizations. As in the elastic case, the predictions of the distinct models for the inelastic photon distribution are similar to small - $x$ and differ for large - $x$, with the LUXqed prediction being an upper bound for $f^{\textrm{inel}}$. In comparison to $f^{\textrm{el}}_{\gamma}$, the inelastic photon distribution dominantes for small values of $x$, which is expected since the probability of proton dissociation  when it emits a  photon with  large virtuality is very high. The results presented in  Fig. \ref{Fig:fgamma} (right panel) indicate that the treatment of the non - exclusive processes will also be impacted by the current uncertainty on the modelling of the inelastic photon distribution. In the next Section, we will estimate the impact of these models for the elastic and inelastic photon distributions on the invariant mass distributions for the dimuon production.


\section{Results}
\label{sec:res}

\begin{figure}[t]
\centering
\includegraphics[width=1.\textwidth]{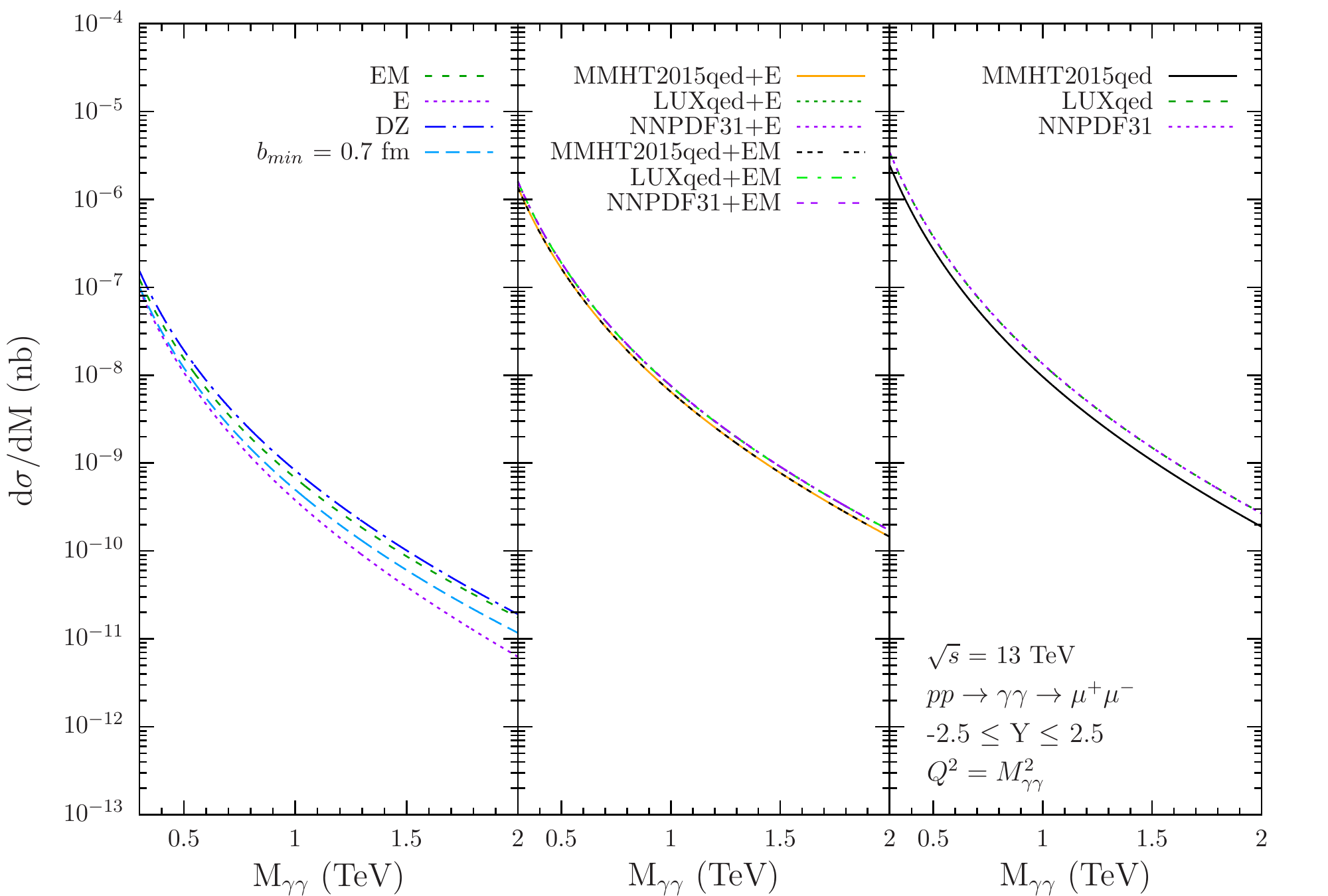}
\caption{\label{Fig:compALL} Invariant mass distributions for the elastic (left), semi - elastic (central) and inelastic (right) dimuon production by $\gamma \gamma$ interactions in $pp$ collisions at $\sqrt{s} = 13$ TeV. Distinct lines represent different combinations of models for the elastic and inelastic photon distributions.  }
\end{figure}

Initially, let's estimate the invariant mass distribution for the dimuon production by $\gamma \gamma$ interactions in $pp$ collisions considering the distinct models for the elastic and inelastic photon distributions discussed in the previous Section. The differential cross section for the production of a $\mu^+ \mu^-$ pair with invariant mass $M_{\gamma\gamma}$ at the rapidity $Y$  is given by 
\begin{eqnarray}
\frac{d \sigma^i}{d M_{\gamma\gamma}} &=& 2 M_{\gamma\gamma} \int \dif Y \,\, \frac{\partial^{2} {\cal{L}}^i_{eff}}{\partial M_{\gamma\gamma}^{2}\partial Y} \cdot \hat{\sigma}_{\gamma\gamma\to \mu^+ \mu^-}(M_{\gamma\gamma}^{2}=x_{1}x_{2}s) , \\
\end{eqnarray}
where $i$ denotes a elastic, semi - elastic or inelastic process and the differential effective photon - photon luminosity is given by the product of the associated  photon distributions evaluated for a given  momentum fraction  $x_{i}$, which can be expressed by $x_{1,2} = \left(  {M_{\gamma\gamma}}/{\sqrt{s}} \right)\exp(\pm Y)$.
In our analysis the hard scale $Q^2$ for the inelastic photon distributions will be assumed as being $M_{\gamma\gamma}^2$. The cross section for the subprocess $\gamma\gamma\to \mu^+ \mu^-$  is given by the Breit-Wheeler formula for the dilepton production via gamma-gamma fusion:
\begin{eqnarray}\nonumber
\hat{\sigma}_{\gamma\gamma\to\mu^{+}\mu^{-}} (M_{\gamma\gamma}^{2}) = \frac{4\pi \alpha^{2}}{M_{\gamma\gamma}^{2}} \left\{ 2 \ln \left[ \frac{1}{\sqrt{\eta_{\mu}}} \left( 1 + \sqrt{\frac{1}{\eta_{\mu}} - 1} \right) \right] \left[ 1 + \eta_{\mu} - \frac{\eta_{\mu}^{2}}{2} \right] - \left( 1 + \eta_{\mu} \right) \sqrt{1 - \eta_{\mu}} \right\}, \\
\label{xsecMUMU}
\end{eqnarray}
where $\eta_{\mu} = 4m_{\mu}^{2}/M_{\gamma\gamma}^{2}$ and the muon mass is taken as $m_{\mu}=0.105658$~GeV \cite{pdg}.

\begin{figure}[t]
\centering
\includegraphics[width=.49\textwidth]{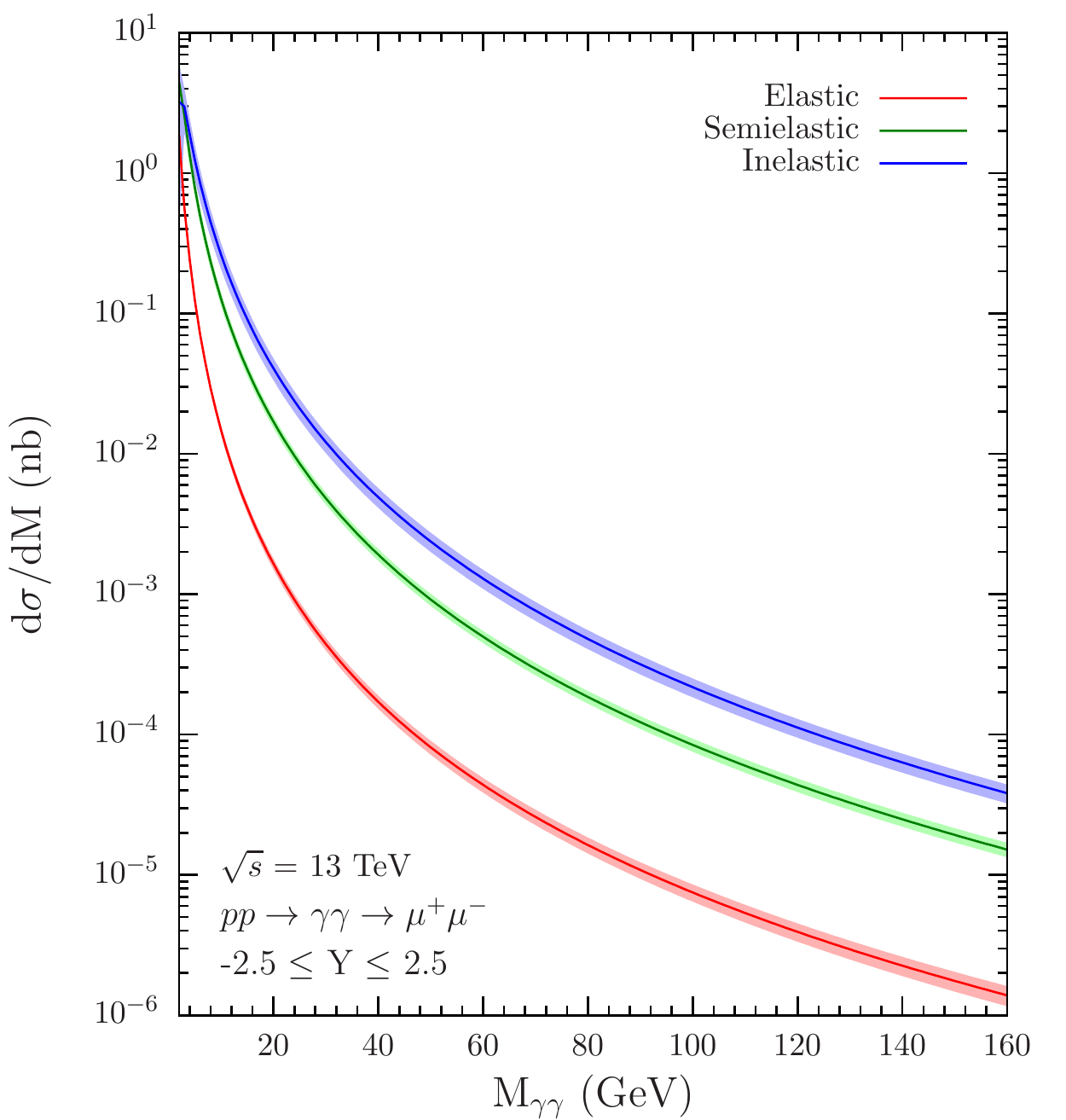}
\includegraphics[width=.49\textwidth]{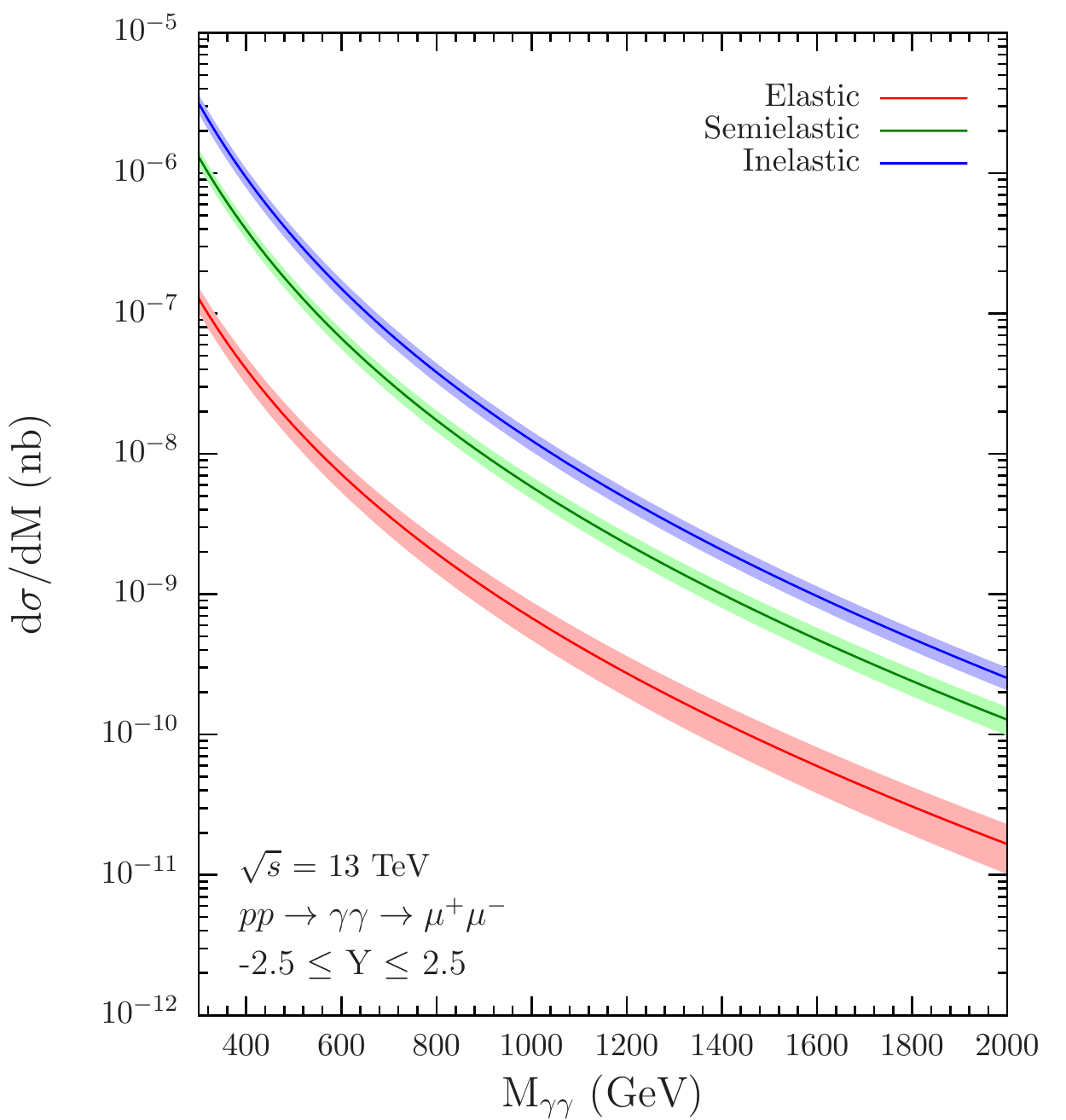}
\caption{\label{Fig:compALL_ave} Invariant mass distributions for the  dimuon production by $\gamma \gamma$ interactions in $pp$ collisions at $\sqrt{s} = 13$ TeV considering two distinct ranges of $M_{\gamma \gamma}$. The solid lines are the average values for the predictions and the band represent the one standard deviation uncertainty based on the different predictions.}
\end{figure}

In Fig. \ref{Fig:compALL} we present our predictions for the elastic (left panel), semi - elastic (central panel) and inelastic (right panel) production of a dimuon pair in $pp$ collisions at $\sqrt{s} = 13$ TeV considering the rapidity range covered by a typical central detector ($|Y| \le 2.5$). The differences are mainly on the normalization, with the shape of the distributions being similar. One has that for the invariant mass range considered, the inelastic production dominates, followed by the semi - elastic one. For the inelastic production, the MMHT2015qed prediction can be considered a lower bound. As expected from the analysis  performed in the previous Section, ${d \sigma^i}/{d M_{\gamma\gamma}}$ is larger for smaller values of $M_{\gamma\gamma}$ and the predictions for large invariant masses are strongly dependent of the model assumed to describe the photon distributions, especially in the elastic case. In order to demonstrate more clearly these conclusions, in Fig. \ref{Fig:compALL_ave} we present the average values of the predictions  with one standard deviation uncertainty,  computed as the absolute deviations summed in quadrature, for the different models considering  the dimuon production with small (left panel) and large (right panel) invariant masses. Our results indicate that a precise treatment of the elastic photon distribution is fundamental to derive a more precise prediction of the elastic cross section for large $M_{\gamma\gamma}$. For the non - exclusive processes, the predictions are also impacted by the distinct models for $f^{\textrm{inel}}$.   These uncertainties may be possibly reduced with the upcoming data from the LHC experiments at this mass range measured with forward detectors.

In what follows we will calculate the ratio between the different contributions for the dimuon production and estimate its dependence on the invariant mass. As pointed out in the Introduction, such study is motivated by the analyzes performed by the CMS Collaboration in the Refs.~\cite{Chatrchyan:2013foa,Khachatryan:2016mud}, where an estimate for the non - exclusive processes was obtained by rescaling the elastic prediction by a data-driven constant factor. Another motivation is to calculate the relative contribution of the non - exclusive processes and estimate the current theoretical uncertainty present in the predictions. In particular, we will estimate the following fractions:
\begin{eqnarray}
F_1 = \frac{\frac{d\sigma^{\textrm{el}}}{dM_{\gamma\gamma}} + \frac{d\sigma^{\textrm{semi}}}{dM_{\gamma\gamma}} + \frac{d\sigma^{\textrm{inel}}}{dM_{\gamma\gamma}}}{\frac{d\sigma^{\textrm{el}}}{dM_{\gamma\gamma}}} & \,\,\,\,\,\, \mbox{and} \,\,\,\,\,\, &
F_2 =  \frac{\frac{d\sigma^{\textrm{el}}}{dM_{\gamma\gamma}} + \frac{d\sigma^{\textrm{semi}}}{dM_{\gamma\gamma}}  }{\frac{d\sigma^{\textrm{el}}}{dM_{\gamma\gamma}}} \,\,.
\label{F_PPS}
\end{eqnarray}
A possible measurement of these factors could provide a deeper insight in the non-exclusive contributions in two - photon interactions. Such experimental results are very challenging to be obtained with central detectors in the LHC experiments, since this difficulty lays on the proper selection to amount the exclusive events. Besides, mostly of the data will cover all the contributions in case no information about the outgoing protons is available. Luckily, the LHC experiments are setting up forward detectors to collect this information, such as PPS of the CMS Collaboration and AFP of the ATLAS Collaboration. These forward detectors are meant to collect the intact outgoing protons from the elastic process at the interacting point. Such protons will scatter in very small angles -- typically $\sim$1~milliradians \cite{review_forward} -- since the proton loose a very small amount of their initial energies. By selecting these protons, the experiments will be able to collect the elastic events with high precision, which is the essential contribution to account for the fraction and to constrain the description of the elastic distribution.

\begin{figure}[t]
\centering
\includegraphics[width=.49\textwidth]{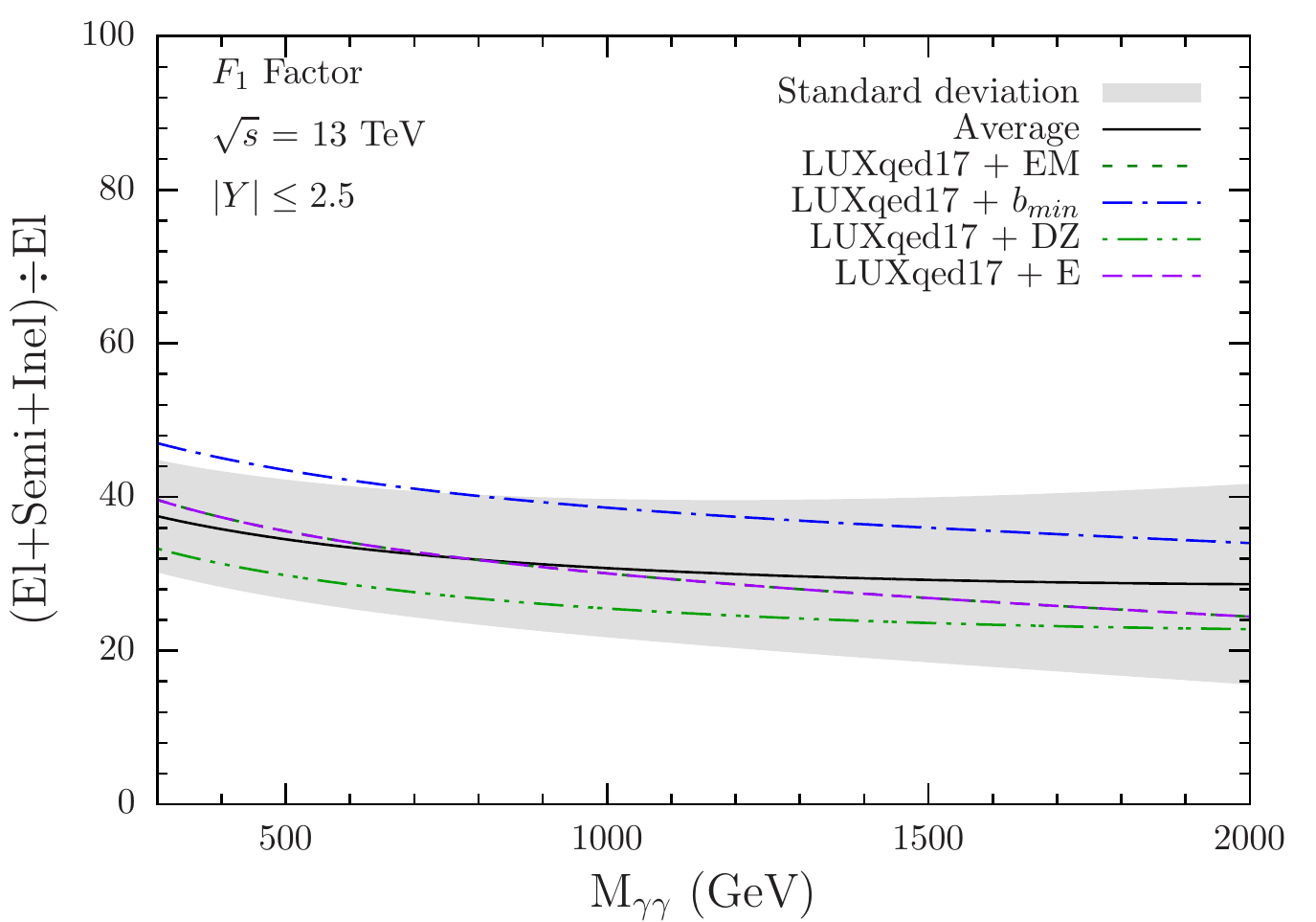}
\includegraphics[width=.49\textwidth]{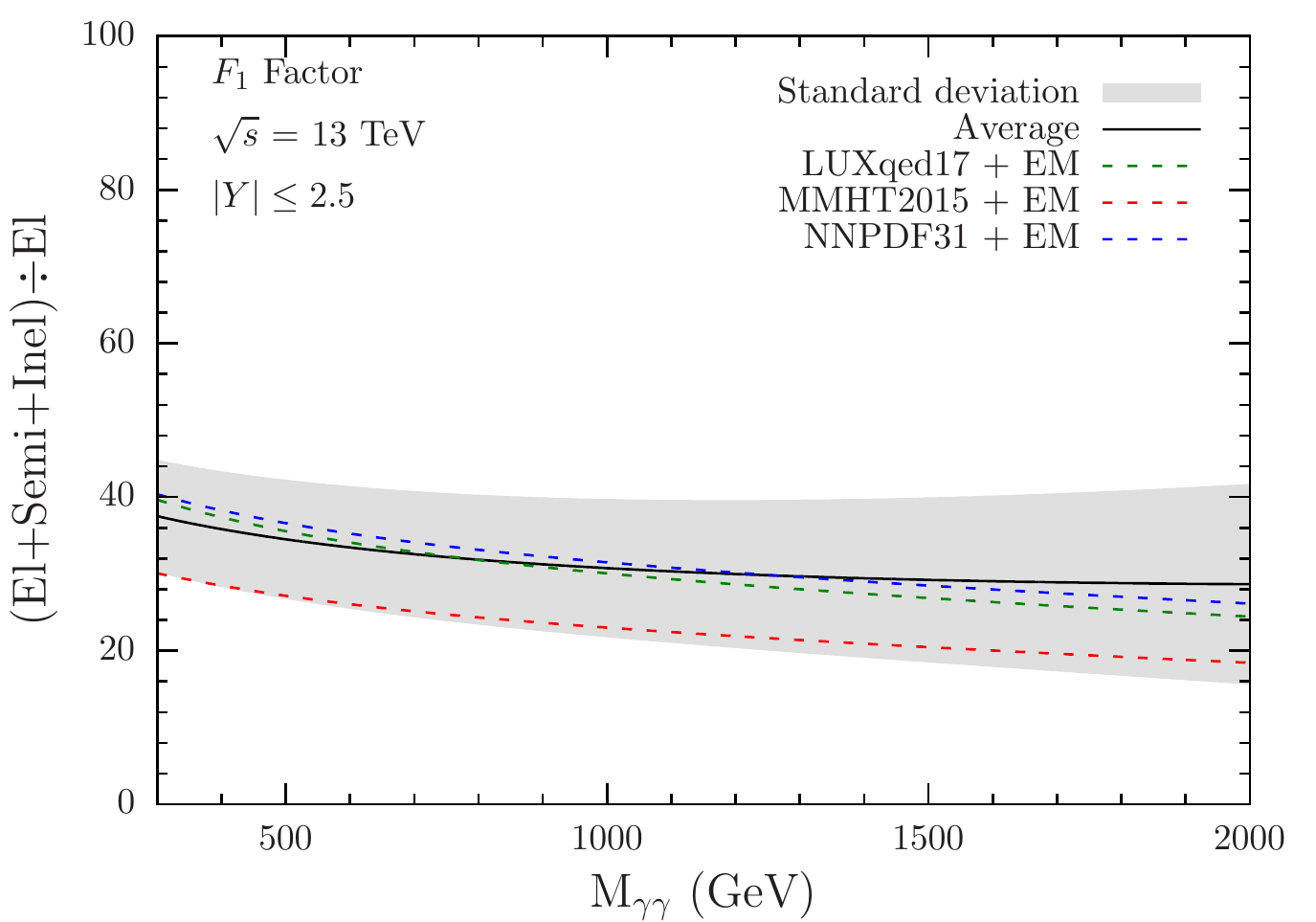}

\caption{\label{Fig:ratioALLel}  Dependence on the invariant dimuon mass of the fraction $F_1$ for  different models of the elastic (left panel) and inelastic (right panel) photon distributions.}
\end{figure}

\begin{figure}[t]
\centering
\includegraphics[width=.49\textwidth]{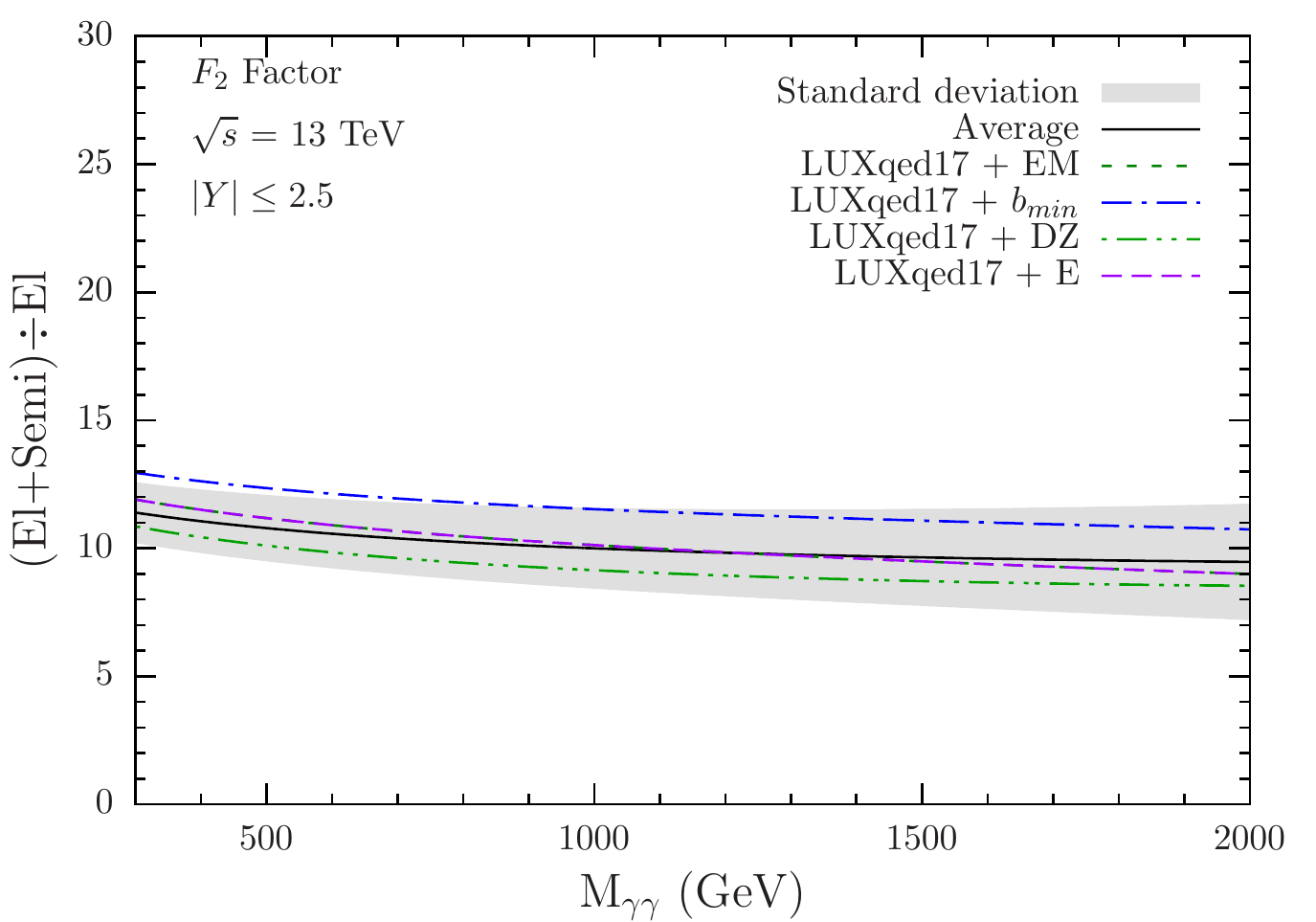}
\includegraphics[width=.49\textwidth]{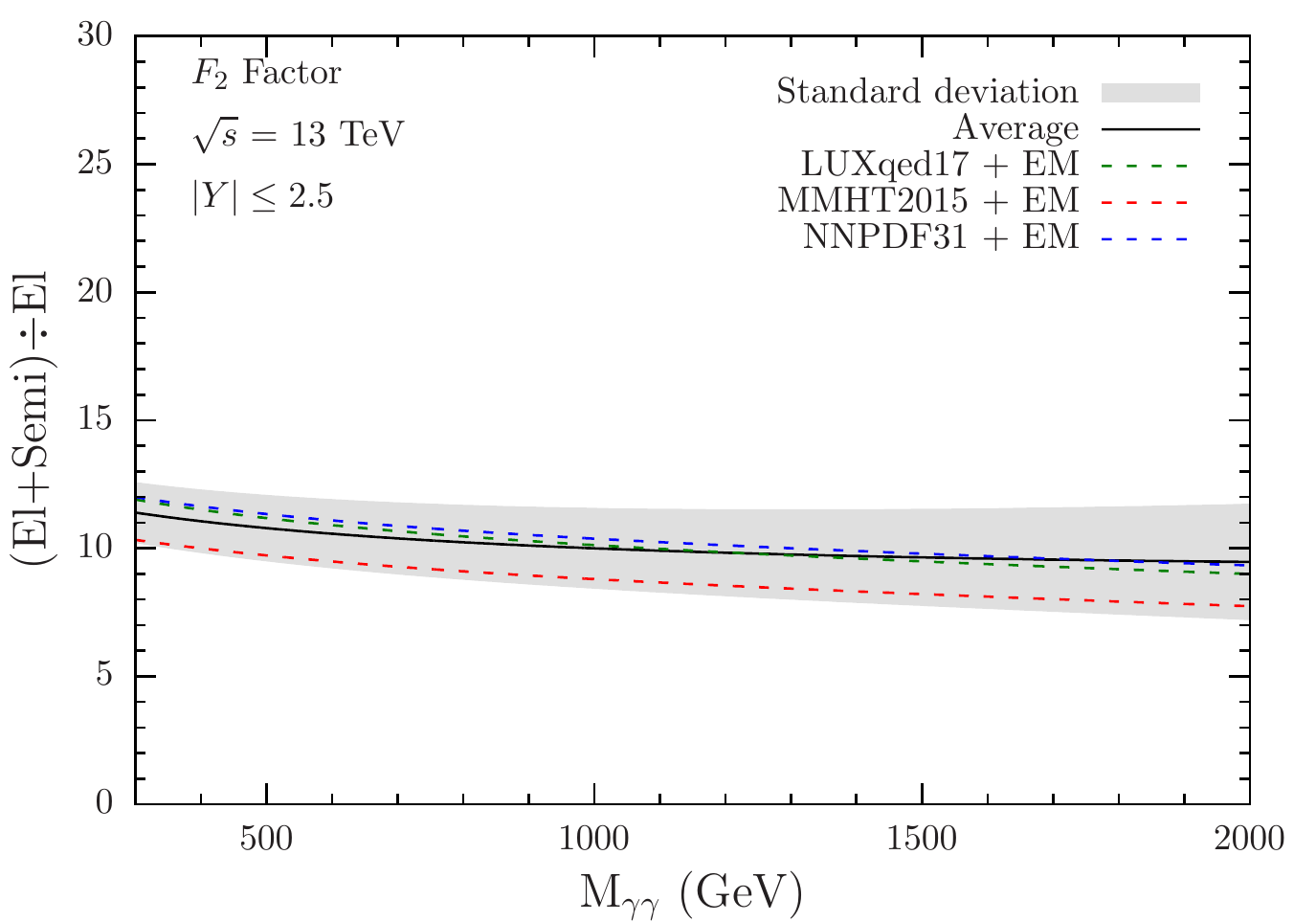}

\caption{\label{Fig:rationonexel} Dependence on the invariant dimuon mass of the fraction $F_2$ for  different models of the elastic (left panel) and inelastic (right panel) photon distributions.}
\end{figure}

In Fig. \ref{Fig:ratioALLel} we present our predictions for the fraction $F_1$  considering different models for the elastic and inelastic photon distributions. Such ratio is useful if the protons are not tagged in the final state by forward detectors. As expected from the results presented in Figs. \ref{Fig:compALL} and \ref{Fig:compALL_ave}, the ratio is much larger than one, which demonstrates that the dimuons with large invariant masses are dominantly produced by non - exclusive $\gamma \gamma$ interactions.   One has that the distinct predictions present a mild dependence on $M_{\gamma \gamma}$ in the range considered, with the magnitude of the ratio being dependent on the model considered. In particular, the calculation of the elastic contribution using the pointlike form factor, Eq. (\ref{Eq:point}), implies a larger amount of non - exclusive processes. In constrast, the combination LUXqed17 + EM, which is currently considered the more precise prediction, implies that this ratio is $\approx 35$ for $M_{\gamma \gamma} = 1000$ GeV. The predictions for the fraction $F_2$ are presented in Fig. \ref{Fig:rationonexel}. Such ratio is of interest if only one of the protons in the final state is tagged by the forward detectors. For this case, the ratio assumes values of the order of 11 in the $M_{\gamma \gamma}$ range considered. Again, the ratio is almost constant and its value depends on the model assumed for the photon distributions. It is important to emphasize that we also have performed the analysis for the $W^+ W^-$ production and obtained similar results for the ratios discussed above. Therefore, the results presented in Figs. \ref{Fig:ratioALLel} and \ref{Fig:rationonexel} indicate that the assumption assumed in the CMS studies is a good first approximation.

The results presented in this Section indicate that the contribution of the non - exclusive processes is dominant and that a future measurement of the distinct contributions will be useful to contrain the description of the elastic and inelastic photon distributions.
Both elastic and non-exclusive processes can be well distinguish  when the acoplanarity, $ a = 1 - |\Delta\phi(x^{+}x^{-})|/\pi$ ($x = \mu, \, W$), and transverse momentum balance, $ \Delta p_{T} = |p_{T}(x^{+}) - p_{T}(x^{-})|$, variables are investigated.  Besides, these contributions may be refined by tagging one or both protons in the final state, which will allow to estimate the contribution associated to non - exclusive events. In principle,  one may be able to obtain measurements for single dissociation events by collecting data from central detectors, selecting events without one proton in the forward detectors and within small acoplanarity and $p_{T}$ balance of the produced pair. This measurement can be a way to narrow down the non-exclusive contribution in two - photon interactions.


\section{Summary} 
\label{sec:sum}
Photon - induced processes are becoming increasingly relevant for phenomenology at the LHC, strongly motivated by the possibility of search for BSM physics and  constrain its different scenarios. In order to derive precise predictions, it is essential to know the relative contribution of the exclusive and non - exclusive processes, which is determined by the photon content of the proton and can  be assessed with experimental data obtained by dedicated forward detectors. In this paper we have reviewed the distinct modellings for the elastic and inelastic photon distributions found in the literature and presented a comparison between its predictions. We have demonstrated that the different models mainly differ in its predictions for  large values of the momentum fraction carried by the photon. We also shown that such uncertainty has direct impact on the predictions for the production of pairs with large invariant mass. Our results indicated that a precise treatment of the elastic photon distribution is vital to estimate the contribution of the exclusive production. Moreover, we have  calculated the relative contribution of the non - exclusive processes and estimated the current theoretical uncertainty present in the predictions. The associated fractions present a mild dependence on the invariant mass of the pair in the kinematical range covered by the LHC, which validates the approach used in recent experimental analyzes. Finally, our results indicate that  a future experimental analysis of the exclusive and non - exclusive processes will be useful to constrain the description of the photon content of proton.

\acknowledgments
This work was partially financed by the Brazilian funding agencies CNPq, CAPES, FAPERGS (process number 17/2551-0001131-7), and INCT-FNA (process number 464898/2014-5).


\end{document}